\documentclass[aps,pra]{revtex4}
\usepackage{graphicx}

\begin{document}

\title{Multiparty quantum secret sharing of secure direct communication
\thanks{Corresponding author's email address: zhangzj@wipm.ac.cn}}
\author{Zhan-jun Zhang$^{1,2}$ \\
{\normalsize $^1$ Wuhan Institute of Physics and Mathematics,
Chinese Academy of Sciences, Wuhan 430071, China} \\
{\normalsize $^2$ School of Physics \& Material Science, Anhui University, Hefei 230039, China} \\
{\normalsize Email: zhangzj@wipm.ac.cn}}

\date{\today}
\maketitle

\begin{minipage}{420pt}
Based on the two-step protocol [Phys. Rev. A68(03)042317], we
propose a $(n,n)$-threshold multiparty quantum secret sharing
protocol of secure direct communication. In our protocol only all
the sharers collaborate can the sender's secure direct
communication message be extracted. We show a variant version of
this protocol based on the variant two-step protocol. This variant
version can considerably reduce the realization difficulty in
experiment. In contrast to the use of multi-particle GHZ states in
the case that the sharer number is larger than 3, the use and
identification of Bell states are enough in our two protocols
disregarding completely the sharer number, hence, our protocols are more feasible in technique.\\

\end{minipage}\\

Quantum key distribution (QKD) is an ingenious application of
quantum mechanics, in which two remote legitimate users (Alice and
Bob) establish a shared secret key through the transmission of
quantum signals and thereafter they can use this key to encrypt
the secret messages or decrypt the encrypted messages. With
quantum mechanics, other cryptographic task can be realized, such
as quantum secret sharing (QSS). QSS is a generalization of
classical secret sharing[1,2] to a quantum scenario [3], where the
sender's secret message is distributed via quantum mechanical
method among $n$ sharers in such a way that only all the sharers
cooperate can they recover the secret message. The QSS is likely
to play a key role in protecting secret quantum information, e.g.,
in secure operations of distributed quantum computation, sharing
difficult-to-construct ancilla states and joint sharing of quantum
money [6], and so on. Hence, after the pioneering QSS work
proposed by Hillery {\it et al.} using three-particle and
four-particle GHZ states [3], this kind of works on QSS attracted
a great deal of attentions in both theoretical and experimental
aspects [4-16,27]. All these works [3-16] can be divided into two
kinds, one only deals with the QSS of classical messages (i.e.,
bits)[5-6,8-11,13-14,27], or only deals with the QSS of quantum
information [4,7,12,15-16] where the secret is an arbitrary
unknown state in a qubit; and the other [3] studies both, that is,
deals with QSS of classical messages and QSS of quantum
information simultaneously. In all these works[3-16,27], when the
party number is greater than 3, multi-particle GHZ states are used
with only two exceptions[13,27]. Note that, so far all these works
dealing with the QSS of classical messages (bits) allow {\it in
essence} a sender to establish a joint key with receivers.

On the other hand, a novel concept, quantum secure direct
communication (QSDC) has been proposed and pursued
recently[17-20]. Different from QKD whose object is to distribute
a common key between the two remote legitimate users of
communication, QSDC can transit the secret messages directly
without creating first a key to encrypt them. Hence, QSDC may be
important in some applications which have been shown by Bostrom
{\it et al.}[18] and Deng {\it et al.}[19-20].

Our one basic purpose in this paper is to generalize the concept
of usual multiparty quantum secret sharing of key to the
counterpart concept of multiparty quantum secret sharing of secure
direct communication. As same as the advantages of QSDCs over
QKDs, the advantages of such a generalization are intuitive. Based
on a QSDC protocol (i.e., the two-step protocol) proposed by Deng
{\it et al.}[19] using Einstein-Podolsky-Rosen (EPR) photon pair
block, in this paper we will propose a multiparty quantum secret
sharing protocol of secure direct communication. In our this
protocol, only all the sharers collaborate can the secure direct
communication message be obtained by all the sharers. Moreover, we
will show a variant version of this protocol based on the variant
two-step protocol we will propose. This variant version can
considerably reduce the realization difficulty in experiment. In
contrast to the use of multi-particle GHZ states in the case that
the sharer number is larger than 3, the use and identification of
Bell states are enough in our two protocols disregarding
completely the sharer number, hence, our protocols are more
feasible in technique. One will see these later.

Now let us turn to our multiparty quantum secret sharing (QSS)
protocol of secure direct communication. For convenience, let us
first describe a three-party protocol. Suppose Alice has a
$(2N)$-bit secret message. She wants to send this secret message
to two distant sharers, Bob and Charlie. One of them, Bob or
Charlie, is not entirely trusted by Alice, and she knows that if
the two guys coexist, the honest one will keep the dishonest one
from doing any damages. The two sharers can infer the secret
message only by their mutual assistance. Our following three-party
QSS protocol of secure direct communication can achieve this goal
with 7 steps.

(a) All the parties (Alice and the two sharers Bob and Charlie)
agree on that each of the four Bell states can carry a two-bit
classical message, i.e., $\phi^{+}$, $\phi^{-}$, $\psi^{+}$ and
$\psi^{-}$ correspond to 00, 01, 10 and 11, respectively. Here the
four Bell states are defined as follows:
\begin{eqnarray}
\phi^{+} =(|0\rangle|0\rangle +|1\rangle|1\rangle)/\sqrt{2}=
(|H\rangle|H\rangle +|V\rangle|V\rangle)/\sqrt{2},
\end{eqnarray}
\begin{eqnarray}
\phi^{-} =(|0\rangle|0\rangle -|1\rangle|1\rangle)/\sqrt{2}=
(|H\rangle|V\rangle +|V\rangle|H\rangle)/\sqrt{2},
\end{eqnarray}
\begin{eqnarray}
\psi^{+}=(|1\rangle|0\rangle+|0\rangle|1\rangle)/\sqrt{2}=
(|H\rangle|H\rangle -|V\rangle|V\rangle)/\sqrt{2},
\end{eqnarray}
\begin{eqnarray}
\psi^{-}=(|1\rangle|0\rangle-|0\rangle|1\rangle)/\sqrt{2}= (
|V\rangle|H\rangle-|H\rangle|V\rangle)/\sqrt{2},
\end{eqnarray}
where $|0\rangle$ and $|1\rangle$ are the up and down eigenstates
of $\sigma_z$, $|H\rangle=(|0\rangle + |1\rangle)/\sqrt{2}$ and
$|V\rangle=(|0\rangle - |1\rangle)/\sqrt{2}$ are the up and down
eigenstates of $\sigma_x$. For convenience,
\{$|0\rangle$,$|1\rangle$\} is refereed to as the diagonal basis
and \{$|H\rangle$,$|V\rangle$\} the rectilinear basis hereafter.

(b) First Alice prepares an ordered $N$ EPR photon pair sequence
according to her secret message. Then she prepares $k+j$ EPR
photon pairs each randomly in one of the four Bell states. We call
these photon pairs as checking photon pairs. Alice randomly
inserts each of the $k+j$ checking photon pairs one by one into
the ordered $N$ photon pair sequence and records each position. We
denote the ordered $N+k+j$ EPR pair sequence with $[(X_1, Y_1),
(X_2, Y_2),\dots, (X_{N+k+j}, Y_{N+k+j})]$. Here the subscript
indicates the pair order in the sequence, and $X$ and $Y$
represent the two photons, respectively.

(c) Alice takes one photon from each EPR pair to form an ordered
photon sequence, say, $[Y_1, Y_2, \dots, Y_{N+k+j}]$. It is called
the $Y$ sequence. The remaining EPR partner photons form another
ordered sequence $[X_1, X_2, \dots, X_{N+k+j}]$. It is called the
$X$ sequence. Alice sends the $Y$ sequence to Bob.

(d) After he receives the $Y$ sequence, for each photon in the
sequence Bob randomly chooses a local unitary operation from
$U_1$, $U_2$, $U_3$, $U_4$ and $H$ and performs this operation on
it. After his encryptions (i.e., the local unitary operations),
Bob sends the ordered $Y$ photons to Charlie. Here
\begin{eqnarray}
U_1=|0\rangle\langle0|+|1\rangle\langle1|=I,
\end{eqnarray}
\begin{eqnarray}
U_2=|0\rangle\langle0|-|1\rangle\langle1|=\sigma_z,
\end{eqnarray}
\begin{eqnarray}
U_3=|0\rangle\langle1|+|1\rangle\langle0|=\sigma_x,
\end{eqnarray}
\begin{eqnarray}
U_4=|0\rangle\langle1|-|1\rangle\langle0|=i\sigma_y,
\end{eqnarray}
\begin{eqnarray}
H=(|0\rangle\langle0|-|1\rangle\langle1|+|0\rangle\langle1|+|1\rangle\langle0|)/\sqrt{2}=
(U_2+U_3)/\sqrt{2}=(\sigma_z+\sigma_x)/\sqrt{2},
\end{eqnarray}
where $H$ is the so-called Hadamard operator. Define
\begin{eqnarray}
\xi \equiv (\phi^-+\psi^+)/\sqrt{2}=
(|H\rangle|0\rangle-|V\rangle|1\rangle)/\sqrt{2}
=(|0\rangle|H\rangle+|1\rangle|V\rangle)/\sqrt{2},
\end{eqnarray}
\begin{eqnarray}
\eta\equiv
(\phi^+-\psi^-)/\sqrt{2}=(|V\rangle|0\rangle+|H\rangle|1\rangle)/\sqrt{2}=(|0\rangle|H\rangle-|1\rangle|V\rangle)/\sqrt{2},
\end{eqnarray}
\begin{eqnarray}
\chi\equiv
(\psi^-+\phi^+)/\sqrt{2}=(|H\rangle|0\rangle-|V\rangle|1\rangle)/\sqrt{2}=(|0\rangle|V\rangle+|1\rangle|H\rangle)/\sqrt{2},
\end{eqnarray}
\begin{eqnarray}
\zeta\equiv(\psi^+-\phi^-)/\sqrt{2}=(-|V\rangle|0\rangle+|H\rangle|1\rangle)/\sqrt{2}
=(-|0\rangle|V\rangle+|1\rangle|H\rangle)/\sqrt{2}.
\end{eqnarray}
For convenience, \{$\phi^{+}$, $\phi^{-}$, $\psi^{+}$,
$\psi^{-}$\} is refereed to as the Bell state set and \{$\xi$,
$\eta$, $\chi$, $\zeta$\} the rotation state set hereafter. The
nice feature of the $U_i (i=1,4)$ operation is that it transforms
the state in both state sets (cf., Table 1 and Table 2). In
contrast, the nice feature of $H$ is that it can realize the state
transformation between the two sets (cf., Table 1 and Table 2).
The purpose of choosing a set of five unitary operations is to
protect the channel between Alice and Bob from Charlie's
interception. For example, if Bob chooses randomly a unitary
operation from only $U_i (i=1,4)$, Charlie could intercept the
channel between Alice and Bob and replaces the $Y$ sequence by a
fake one. Then when Bob sends the encrypted fake $Y$ sequence to
Charlie, Charlie can know Bob's encryptions (i.e., his $U_i
(i=1,4)$ operations) by Bell-state measurements on his fake pair
sequence. Since Charlie has already had all the information about
Bob's encryptions, he can perform the same encryptions as Bob did
on his intercepted $Y$ sequence and accordingly he can readily
retrieve the complete message without Bob's help.

\begin{center}
Table 1 Results of four Bell states after the $U_1^Y$, $U_2^Y$,
$U_3^Y$, $U_4^Y$ and
$H_Y$ unitary operations. See text for details. \\
\begin{tabular}{ccccccccc}\hline
&& $\phi^+_{XY}$ && $\phi^-_{XY}$ && $\psi^+_{XY}$ &&$\psi^-_{XY}$\\
\hline
$U_1^Y$ && $\phi^+_{XY}$ &&  $\phi^-_{XY}$ && $\psi^+_{XY}$ &&  $\psi^-_{XY}$\\
$U_2^Y$ && $\phi^-_{XY}$ &&  $\phi^+_{XY}$ && $\psi^-_{XY}$ &&  $\psi^+_{XY}$\\
$U_3^Y$ && $\psi^+_{XY}$ && -$\psi^-_{XY}$ && $\phi^+_{XY}$ && -$\phi^-_{XY}$\\
$U_4^Y$ && $\psi^-_{XY}$ && -$\psi^+_{XY}$ && $\phi^-_{XY}$ && -$\phi^+_{XY}$ \\
$H_Y$ && $\xi_{XY}$ && $\eta_{XY}$ &&$\chi_{XY}$ &&$\zeta_{XY}$\\
\hline
\end{tabular}\\
\end{center}

\vskip 0.5cm
\begin{center}
Table 2 Results of the $\xi_{XY}$, $\eta_{XY}$, $\chi_{XY}$ and
$\zeta_{XY}$ states after the $U_1^Y$, $U_2^Y$, $U_3^Y$, $U_4^Y$
and $H_Y$ unitary operations. See text for details. \\
\begin{tabular}{ccccccccc}\hline
       && $ \xi_{XY}  $ && $ \eta_{XY} $ && $ \chi_{XY} $ && $ \zeta_{XY}$\\\hline
$U_1^Y$&& $ \xi_{XY}  $ && $ \eta_{XY} $ && $ \chi_{XY} $ && $ \zeta_{XY}$\\
$U_2^Y$&& $ \chi_{XY} $ && $-\zeta_{XY}$ && $ \xi_{XY}  $ && $-\eta_{XY} $\\
$U_3^Y$&& $ \eta_{XY} $ && $ \xi_{XY}  $ && $ \zeta_{XY}$ && $ \chi_{XY} $\\
$U_4^Y$&& $-\zeta_{XY}$ && $ \chi_{XY} $ && $-\eta_{XY} $ && $\xi_{XY}  $\\
$H_Y$&& $\phi^+_{XY}$ && $\phi^-_{XY}$ && $\psi^+_{XY}$&& $\psi^-_{XY}$\\
\hline
\end{tabular}
\end{center}
\vskip 1cm

(e) After confirming that Bob has received an ordered $Y$
sequence, Alice publicly announces the positions of $k$ checking
photon pairs. For each of these $k$ checking photon pairs, Alice
first lets Charlie measure the corresponding checking photon in
his lab by using the diagonal basis or the rectilinear basis and
publish his measurement outcome. Alice's choice of letting Charlie
use which basis is completely random. Then Alice lets Bob publish
which unitary operation he has performed on the checking photon
when it passes by him. If Bob's unitary operation is $H$, then
Alice measures the corresponding checking photon in her lab by
using the different basis from the one she let Charlie use.
Otherwise, she measures the photon using the same basis as Charlie
used. By the way, if only one basis is used, then Eve can steal
secret information without being detected. Since only Alice knows
the initial state of each checking photon pair, after all her
measurements, she can determine the error rate according to the
correlations (cf., equations 1-4 and 10-13). If the error rate
exceeds the threshold, the process is aborted. Otherwise, the
process continues and Alice sends the retained $N+j$ ordered
sequence to Charlie.

(f) After Charlie receives the photons, if Bob and Charlie
collaborate, both Bob and Charlie can obtain a raw secret message
including Alice's secret message and additional $2j$ useless bits.
The detailed procedure is as follows. For each photon pair, if
Bob's unitary operation is $U_i (i=1,4)$, then Bob and Charlie
directly perform a Bell-state measurement on the photon pair. If
Bob's unitary operation is $H$, then they first perform a
Hardamard operation on the $Y$ photon and then perform a
Bell-state measurement on the photon pair. According to their
measurement outcome and Bob's exact unitary operation they can
deduce the two secret bits Alice imposed on this photon pair. For
example, without loss of generality, suppose their outcome is
$\psi^-$. If Bob's unitary operation is $U_3$, then according to
Table 1 they can infer that the secret bits Alice imposed on this
photon pair is '01' corresponding to the initial state $\phi^-$;
if Bob's unitary operation is $H$, they can infer that the secret
bits Alice imposed on this photon pair is '11' corresponding to
the initial state $\psi^-$. On the other hand, if Bob and Charlie
do not collaborate, then neither of them can  get access to
Alice's secret message with 100\% certainty.

(g) Alice publicly announces the initial states and positions of
the retained $j$ checking pairs for Bob and Charlie to extract her
exact secret message from the raw secret message and to check
whether the $N+j$ X photons travelling from Alice site to
Charlie's site have been attacked, which is called message
authentification. Even if the photons are attacked, the
eavesdropper Eve can not get access to any useful information but
interrupt the transmissions quantum channel between Alice and the
last receiver (Charlie).

So far we have proposed the three-party QSS protocol of secure
direct communication based on Deng {\it et al.}'s two-step QSDC
protocol[19] by using the EPR photon pair block. Its security
proof is similar to those in Refs.[18-19,21-23] with entangled
photons. The proof of our this protocol is based on the security
for the transmission of the Y sequence. This is very similar to
the security for the transmission of the C sequence in the
two-step protocol. If Alice, Bob and Charlie could not detect
eavesdropper Eve in the transmission of the Y sequence, the
eavesdropper would adopt the intercept-resend strategy to
eavesdrop the quantum channel and easily reads out the secure
direct communication message. In fact, in our protocol Alice, Bob
and Charlie can know whether the transmission of the Y sequence is
secure or not by using the checking procedure based on the qubit
correlation of each Bell state and Eve will be found out if she
eavesdrops the quantum channel. Moreover, even in a worse case
that there is an insider (Bob or Charlie) who wants to occupy
independently the secure direct communication without another
sharer's help and thereby eavesdrops the quantum line, this
insider's eavesdropping can be also detected by using the
detecting method. Once the security for the transmission of the Y
sequence is not ensured, Alice will abort the transmission of the
X sequence, thereby no information will be leaked to the
eavesdropper. As same as in the two-step protocol, the
transmission and the security check of the Y sequence in our this
protocol is also very similar to the procedure in BBM92 QKD
protocol[26], where one qubit in an EPR pair is sent to a party
and another is sent to another party. Here the X sequence is
retained securely in Alice's lab and Eve (or an insider) can not
get access to it at all before Alice's eavesdropping detection.
Hence, the security of transmission of the Y sequence can be
simply reduced to the security of the BBM92 QKD protocol. The
proof of security for BBM92 in ideal and practical conditions has
been given so far[21,23]. Hence our this protocol is also
unconditionally secure.

Now let us generalize the three-party QSS protocol of secure
direct communication to a $n$-party ($n\geq 4)$ QSS one. Suppose
that Alice is the message sender who would like to send a secret
massage to Bob, Charlie, Dick, \dots, York and Zach (there is
totally $n-1$ sharers). Alice does not trust anyone but the sharer
entirety. She hopes that, only the sharer entirety cooperates can
her secure direct communication message be obtained by the
entirety, otherwise, no one of the entirety can get access to her
secret message with 100\%. Our $n$-party ($n\geq 4)$ QSS  protocol
of secure direct communication can achieve this goal. The first
four step and the last step of the $n$-party ($n\geq 4)$ QSS
protocol of secure direct communication is the same as these in
the three-party QSS protocol. Therefore, we describe the $n$-party
($n\geq 4)$ QSS protocol from the step 5.

(V) After receiving the Y sequence, Charlie encrypts it in the
same way as Bob and then sends it to the next sharer, say, Dick.
Similar procedures are repeated until the $(n-1)$th sharer Zach
receives the Y sequence.

(VI) After confirming that Zach has received the ordered $Y$
sequence, Alice publicly announces the positions of $k$ checking
photon pairs. For each of these $k$ checking photon pairs, Alice
first lets Zach measure the corresponding checking photon in his
lab by using either the diagonal basis or the rectilinear basis
and publish his measurement outcome. Alice's choice of letting
Zach use which basis is completely random. Then Alice lets all
other sharers without Zach publish which unitary operations they
have performed on the checking photon when it passes by them.
Since only Alice knows the initial state of each checking photon
pair, using Table 1 and Table 2 she can works out its final state
before measurements. If the final state belongs to the rotation
state set, then Alice measures the corresponding checking photon
in her lab by using the different basis from the one she has let
Zach use. Otherwise, she measures the photon using the same basis
as she has let Zach use. Similar procedures are repeated until all
actions including Zach's measurements and publishes, all other
sharers' publishes of unitary operations and Alice's unitary
operations and measurements are completed for the $k$ checking
photon pairs. Then Alice can determine the error rate in terms of
the correlations (cf., equations 1-4 and 10-13). If the error rate
exceeds the threshold, the process is aborted. Otherwise, the
process continues and Alice sends the retained $(N+j)$ ordered
sequence to Zach.

(VII) After Zach receives the photons, if all the sharers
collaborate, then they can obtain a raw secret message including
Alice's secret message and additional $2j$ useless bits. The
detailed procedure is as follows. For each photon pair, they the
unitary operation $U=U_B^+U_C^+U_D^+ \dots U_Z^+$ on the Y photon,
where $U_B (U_C, U_D, \dots, U_Z$) represents the unitary
operation Bob (Charlie, Dick, \dots, York, Zach) has ever
performed on the Y photon. After this unitary operation, the state
of this photon pair is recovered to its initial state, thereby all
the sharers can take a Bell-state measurement to extract Alice's
two-bit secret information. After all the photon pairs are dealt
with, they can obtain the raw secret message. On the other hand,
if anyone of all the sharers does not collaborate, then none of
them can get access to Alice's secret message with 100\%
certainty.

(VIII) As mentioned previously, the present last step is same as
the counterpart step in the three-party QSS protocol of secure
direct communication. This step ensures all the sharers to extract
Alice's exact secret message from the raw secret message and the
security of transmission of the $(N+j)$ X photons.

So far we have established a $n$-party QSS protocol of secure
direct communication. Its security is the same as the security of
three-party QSS protocol of secure direct communication, which is
also unconditionally secure.

In our multiparty QSS protocol of secure direct communication,
Alice is the preparer of the EPR photon pair sequence and the X
and Y photon sequences are sent from her by two-steps. Such
unilateral transmissions determines that when all sharers extract
Alice's secret message they should perform reversal unitary
operations to recover the state of each relative pair to its
initial state. It is known that the more operations the more
difficulties in experiment. In the following we will take
advantage of the change of transmission to reduce the difficulty
in experiment. One will see this later.  Suppose Alice has a
$2N$-bit secret message. She would like to send a secret massage
to Bob, Charlie, Dick, \dots, York and Zach (there is totally
$n-1$ sharers). Alice does not trust anyone but the entirety. She
hopes that, only the entirety cooperate can her secure direct
communication message be obtained by the entirety, otherwise, no
one of the entirety can get access to her secret message with
100\%. The variant version of our this $n$-party QSS protocol of
secure direct communication is depicted as follows.

(1) All the parties (Alice and all the sharers Bob, Charlie, Dick,
\dots, York and Zach) agree on that each of the four Bell states
can carry a two-bit classical message, i.e., $\phi^{+}$,
$\phi^{-}$, $\psi^{+}$ and $\psi^{-}$ correspond to 00, 01, 10 and
11, respectively.

(2) Alice lets Zach prepare an ordered $N+k+j$ EPR pair block and
initial states of all the pairs are in $\phi^{+}$. We denote the
ordered $N+k+j$ EPR pair sequence is denoted with $[(X_1, Y_1),
(X_2, Y_2),\dots, (X_{N+k+j}, Y_{N+k+j})]$.

(3) Zach takes one photon from each EPR pair to form an ordered
photon sequence, say, $[Y_1, Y_2, \dots, Y_{N+k+j}]$. It is called
the $Y$ sequence. The remaining EPR partner photons form another
ordered sequence $[X_1, X_2, \dots, X_{N+k+j}]$. It is called the
$X$ sequence. Alice sends the $Y$ sequence to the $(n-2)$th sharer
York.

(4) After he receives the $Y$ sequence, for each photon in the
sequence York randomly chooses a local unitary operation from
$U_1$, $U_2$, $U_3$, $U_4$ and $H$ and performs this operation on
it and then sends it to the $(n-3)$th sharer. The  $(n-3)$th
sharer encrypts the Y sequence in the same way as York and then
sends it to the next sharer, an so on. Similar procedures are
repeated until the first sharer Bob finishes his encryptions. Bob
sends the Y sequence to Alice.

(5) Alice selects randomly $k$ photons and publicly announces
their positions in the Y sequence. Alice lets Zach measure the
partner photons in the X sequence of the selected photons in the Y
sequence either in the diagonal basis or in the rectilinear basis.
To prevent any sharer's intercept-resend attack, for each selected
photon, Alice randomly selects a sharer one by one and let him or
her tell her this sharer's message till she obtains all sharer's
messages. Here Zach's message includes which basis he has used to
measure the partner photon and his measurement outcome, York's
(,\dots, Charlie's, Bob's) message is which unitary operation he
has ever performed on the selected photon. By the way, such random
choice is to protect the quantum channel from any sharer's
interception. Hence, Alice can know via calculation the exact
final state of her selected photon should be after Zach's
measurement on the partner photon, and thereby after her
measurements she can determine the error rate. If the error rate
exceeds the threshold, the process is aborted. Otherwise, the
process continues and Alice performs unitary operations on the
retained photons in the Y sequence to encode a raw secret message
including her secret messages and additional $2j$ useless bits.
Alice sends these encoded photons to Zach.

(6) After Zach receives the photons, if all the sharers
collaborate, then they can obtain a raw secret message including
Alice's secret message and additional $2j$ useless bits. The
detailed procedure is as follows. For each photon pair, they can
work out the exact state before Alice's encoding. If it belongs to
the Bell state set, then they directly take a Bell-state
measurement and thereby they can know the two bits Alice has ever
imposed on this pair. Otherwise, they first perform a Hadamard
operation on the Y photon and take a Bell-state measurement. From
Table 2 they can infer the state of this pair before the Hadamard
operation and thereby they can know the two bits Alice has ever
imposed on the pair. Note that after the change of the
transmission direction the reversal operations are unnecessary
anymore during the raw message extraction. To this sense, we think
this variant version has considerably reduced the realization
difficulty in experiment.

(7) Alice publicly announces positions of the additional $j$ pairs
and her encodings for all the sharers to extract her exact secret
message from the raw secret message and to check whether the
$(N+j)$ Y photons travelling from Alice site to Charlie's site
have been attacked. Even if the photons are attacked, the
eavesdropper Eve can not get access to any useful information but
interrupt the transmissions quantum channel between Alice and
Zach.

So far we have presented the variant version of our multiparty QSS
protocol of secure direct communication. It is based on a variant
two-step protocol we just proposed. Different from the first
multiparty QSS protocol where the sender Alice sends the Y and X
sequence by two steps and her secret message is encoded on the X
sequence, in the variant version it is the last sharer Zach who
sends the Y sequence to fetch the secret message. Such a change of
transmissions ensures the realization difficulty in experiment can
be considerably reduced. As mentioned previously, its security
proof is also very similar these in Refs.[18-19, 21-23] and can be
reduced to the proof of the BBM92 QKD protocol[26]. Hence, it is
also unconditionally secure.

In all our present protocols, the multi-particle GHZ states in all
other existing multiparty QSS schemes are not necessary. Although
in [15] it is claimed that only Bell states are needed, the
identification of multi-particle GHZ state is still necessary. In
our multiparty SSQI protocol, only the use and identification of
Bell states are needed. Till now to our knowledge only five-photon
GHZ state is realized in experiment[24], more-photon GHZ state is
still desired in experiment. Hence, in the case when the sharer
number is large enough, the usual QSS schemes can only be possible
in theory. In contrast, for any large number of sharers, our
protocols can securely work provided the use and identification of
Bell states are possible. Hence, the present multiparty QSS
protocols are more feasible in technique[25].

To summarize, in this paper based on the two-step protocol we have
presented a multiparty quantum secret sharing protocol of secure
direct communication. In our protocol only all the sharers
collaborate can the sender's secure direct communication messages
be extracted. We show a variant version of this protocol based on
the variant two-step protocol. This variant version can
considerably reduce the realization difficulty in experiment. In
contrast to the use of multi-particle GHZ states in the case that
the sharer number is larger than 3, the use and identification of
Bell states are enough in our two protocols disregarding
completely the sharer number, hence, our protocols are more feasible in technique. \\

\noindent {\bf Acknowledgement}

This work is supported by the National Natural Science Foundation
of China under Grant No. 10304022. \\

\noindent {\bf References}

\noindent[1]  B. Schneier, Applied Cryptography (Wiley, New York,
1996) p. 70.

\noindent[2] J. Gruska,  Foundations of Computing (Thomson
Computer Press, London, 1997) p. 504.

\noindent[3] M. Hillery, V. Buzk,  and A. Berthiaume, Phys. Rev. A
{\bf 59}, 1829 (1999).

\noindent[4] R. Cleve, D. Gottesman,  and H. K. Lo, Phys. Rev.
Lett. {\bf 83}, 648 (1999).

\noindent[5] A. Karlsson, M. Koashi, and N. Imoto, Phys. Rev. A
{\bf 59}, 162  (1999).

\noindent[6] D. Gottesman,  Phys. Rev. A {\bf 61}, 042311  (2000).

\noindent[7] S. Bandyopadhyay, Phys. Rev. A {\bf 62}, 012308
(2000).

\noindent[8] W. Tittel, H. Zbinden, and N. Gisin, Phys. Rev. A
{\bf 63}, 042301  (2001).

\noindent[9] V. Karimipour and A. Bahraminasab, Phys. Rev. A {\bf
65}, 042320 (2002).

\noindent[10] H. F. Chau, Phys. Rev. A {\bf 66}, 060302  (2002).

\noindent[11] S. Bagherinezhad and V. Karimipour, Phys. Rev. A
{\bf 67}, 044302  (2003).

\noindent[12] L. Y. Hsu, Phys. Rev. A {\bf 68}, 022306 (2003).

\noindent[13] G. P. Guo and G. C. Guo, Phys. Lett. A {\bf 310},
 247 (2003).

\noindent[14] L. Xiao, G. L. Long, F. G. Deng, and  J. W. Pan,
Phys. Rev. A {\bf 69}, 052307 (2004).

\noindent[15] Y. M. Li, K. S. Zhang, and K. C. Peng, Phys. Lett. A
{\bf 324}, 420  (2004).

\noindent[16] A. M. Lance, T. Symul, W. P. Bowen, B. C. Sanders,
and P. K. Lam,  Phys. Rev. Lett. {\bf 92}, 177903 (2004).

\noindent[17] A. Beige, B. G. Englert, C. Kurtsiefer, and
H.Weinfurter, Acta Phys. Pol. A {\bf101}, 357 (2002).

\noindent[18] K. Bostrom K and F. Felbinger. Phys. Rev. Lett.
{\bf89}, 187902 (2002).

\noindent[19] F. G. Deng, G. L. Long, and  X. S. Liu, Phys. Rev. A
{\bf68},  042317 (2003).

\noindent[20] F. G. Deng and G. L. Long,  Phys. Rev. A {\bf69},
 052319 (2004).

\noindent[21] H. Inamori, L. Rallan,  and V. Verdral, J. Phy. A
{\bf 34}, 6913 (2001).

\noindent[22] G. L. Long and X. S. Liu, Phys. Rev. A {\bf 65},
032302 (2002) .

\noindent[23] E. Waks, A. Zeevi, and Y. Yamamoto, Phys. Rev. A
{\bf 65}, 052310 (2002) .

\noindent[24] Z. Zhao, Y. A. Chen, A. N. Zhang, T. Yang, H. J.
Briegel, and J. W. Pan, Nature (London) {\bf 430}, 54 (2004).

\noindent[25] Y. H. Kim, S. P. Kulik, and Y. Shih, Phys. Rev.
Lett. {\bf 86}, 1370 (2001).

\noindent[26] C. H. Bennett, G. Brassard, and N. D. Mermin, Phys.
Rev. Lett. {\bf68}, 557 (1992).

\noindent[27] Z. J. Zhang, Y. Li, and Z. X. Man, resubmitted to
Phys. Rev. A

\enddocument